\documentclass[12pt,draftclsnofoot, onecolumn]{IEEEtran}

\usepackage{cite}
\usepackage{graphicx}
\usepackage{amsmath, amsthm, amssymb}
\usepackage{algorithm,algorithmicx,algcompatible,algpseudocode}
\usepackage{enumerate}
\usepackage{balance, color}
\usepackage{multirow, verbatim}
\usepackage{url}
\usepackage{subfigure}

\usepackage{tikz}
\usetikzlibrary{arrows, shapes, positioning}
\usepackage{pgfplots}

\theoremstyle{plain} % default
\newtheorem{definition}{Definition}
\newtheorem{theorem}{Theorem}
\newtheorem{proposition}[theorem]{Proposition}  % Proposition having the same counter as theorem [thm]
\newtheorem{lemma}[theorem]{Lemma} % Lemma having the same counter as theorem [thm]
\newtheorem{corollary}[theorem]{Corollary} % Corollary having the same counter as theorem [thm]
\theoremstyle{definition} % different from style default
\newtheorem{example}{Example}

\newcommand{\ie}{{\em i.e.}}
\newcommand{\ea}{{\em et al.}}
\newcommand{\eg}{{\em e.g.}}

\newcommand{\vct}[1]{\ensuremath{\mathbf{#1}}}
\newcommand{\uset}{\ensuremath{\mathcal{U}}}
\newcommand{\pset}{\ensuremath{\mathcal{P}}}
\newcommand{\hset}[1]{\ensuremath{\mathcal{H}_{#1}}}

\newcommand{\wset}[1]{\ensuremath{\mathcal{W}_{#1}}}

\newcommand{\matr}[1]{\mathbf{#1}}

\newcommand{\squishlist}{
 \begin{list}{$\bullet$}
  { \setlength{\itemsep}{0pt}
     \setlength{\parsep}{3pt}
     \setlength{\topsep}{3pt}
     \setlength{\partopsep}{0pt}
     \setlength{\leftmargin}{1.5em}
     \setlength{\labelwidth}{1em}
     \setlength{\labelsep}{0.5em} } }

\newcommand{\squishlisttwo}{
 \begin{list}{--}
  { \setlength{\itemsep}{0pt}
     \setlength{\parsep}{0pt}
    \setlength{\topsep}{0pt}
    \setlength{\partopsep}{0pt}
    \setlength{\leftmargin}{2em}
    \setlength{\labelwidth}{1.5em}
    \setlength{\labelsep}{0.5em} } }

\newcommand{\squishend}{
  \end{list}  }

\begin{document}

%--------------- Title -----------------------------------------------------
\title{Network Codes for Real-Time Applications}

\author{Anh~Le,~\IEEEmembership{Member,~IEEE,}
	Arash~S.~Tehrani,~\IEEEmembership{Member,~IEEE,}
	Alexandros~G.~Dimakis,~\IEEEmembership{Member,~IEEE,}
	and~Athina~Markopoulou,~\IEEEmembership{Senior Member,~IEEE}%
\IEEEcompsocitemizethanks{\IEEEcompsocthanksitem A. Le and A. Markopoulou are with the Department of Electrical Engineering and Computer Science,
University of California, Irvine, CA, 92697.\protect\\
E-mail: \{anh.le, athina\}@uci.edu
\IEEEcompsocthanksitem A. S. Tehrani is with the Department of Electrical Engineering, University of Southern California, CA, 90089.\protect\\
E-mail: arash.sabertehrani@usc.edu
\IEEEcompsocthanksitem A. G. Dimakis is with the Department of Electrical and Computer Engineering, University of Texas, Austin, TX, 78712.}
}%

%--------------- Abstract -----------------------------------------------------
\IEEEtitleabstractindextext{%
\begin{abstract}
%Broadcast is subject to packet losses due to channel impairments. Network coding has been shown to improve transmission efficiency in loss recovery. Existing literature focuses on designing loss recovery schemes that minimize the completion delay, \ie, the time it takes to recover all losses at all users. In real-time applications, however, complete recovery may suffer because of strict and urgent deadlines. Therefore, for these applications, instantly decodable codes is attractive as it guarantees immediate decoding by the users.

We consider the scenario of  broadcasting for real-time applications and loss recovery via instantly decodable network coding. Past work  focused on minimizing the completion delay, which is not the right objective for real-time applications that have strict deadlines. In this work, we are interested in finding  a code that  is instantly decodable by the maximum number of users. First,  we prove that this problem is NP-Hard  in the general case. Then we consider the practical probabilistic scenario, where users have i.i.d. loss probability and the number of packets is linear or polynomial in the number of users. In this scenario, we provide a polynomial-time (in the number of users) algorithm that finds the optimal coded packet. The proposed algorithm is evaluated using both simulation and real network traces of a real-time Android application. Both results show that the proposed coding scheme significantly outperforms the state-of-the-art baselines: an optimal repetition code and a COPE-like greedy scheme.
\end{abstract}

}

\maketitle

\IEEEdisplaynontitleabstractindextext

%--------------- Introduction --------------------------------------------------
\section{Introduction}
\label{sec:intro}

\IEEEPARstart{B}{roadcasting} data to multiple users is widely used in several wireless applications,
ranging from satellite communications to WiFi networks. 
Wireless  transmissions are subject to packet losses due to channel impairments, such as, wireless fading and interference. 
Previous work has shown that coding  can improve transmission efficiency, throughput, and delay over broadcast erasure channels \cite{idnc-netcod07, idnc-netcod08, idnc-netcod09, idnc-icc10, ec-sigcomm98, ec-ccnc06}. 
Intuitively, the diversity of lost packets across different users creates coding opportunities that can improve various performance metrics.

%\footnote{\textcolor{blue}{[all these are idnc references. how about some erasure coding for loss recovery?]}} -- Added some ec papers

In this work, we are interested in packet recovery for real-time applications, such as, fast-paced multi-player games and live video streaming. Real-time applications have two distinct characteristics: (i) they have strict and urgent deadlines, \ie, a packet is outdated after a short amount of time, and (ii) they can tolerate some losses, \eg, a game client can restore its state in the presence of losses by resyncing periodically \cite{microplay}. 
%
% Guidelines for loss recovery
Although having limited fault tolerance, these applications may suffer significantly from packet losses and lead to poor performance, \eg, jittery game animation and low quality video playback. Hence, it is highly desirable to recover packet losses with very low delay and within a very narrow coding window. Motivated by the above observations, we focus on coding schemes for loss recovery that allows instantaneous decoding, \ie, with zero delay. These coding schemes are also known as Instantly Decodable Network Codes (IDNC).

% Besides the zero delay, benefits of using IDNC include (i) small receiver buffers as the receivers do not need to buffer coded packets, and (ii) low coding complexity as encoding and decoding can be implemented using simple XOR operations. These benefits make IDNC schemes ideal for real-time applications on resource-constrained platforms, such as smartphones.

Previous work on IDNC~\cite{idnc-netcod08, idnc-netcod09, idnc-icc10, idnc-globecom10, idnc-icc11, idnc-pimrc11} focused on minimizing  the completion delay, \ie, the time it takes to recover all the losses at all users. We formulate a different problem that  is more relevant to real-time applications, called {\em Real-Time IDNC}: Consider a source that broadcasts a set of packets, $\mathcal{X}$, to a set of users, $\mathcal{U}$.   Each user, $u \in \mathcal{U}$, wants all packets in $\mathcal{X}$ and already knows a subset of them, $\mathcal{H}_u \subset \mathcal{X}$, for example, through previous transmissions. The goal is to choose one (potentially coded) packet to broadcast from the source, so as to maximize the number of users who can immediately recover one lost packet. This problem is highly relevant in practice, yet -- to the best of our knowledge -- only solved in heuristic ways so far, \eg, see \cite{idnc-ton08, idnc-netcod08}. Our main contributions are the following:

\squishlist
\item We show that Real-Time IDNC is NP-hard. To do so, we first map {Real-Time IDNC} to the Maximum Clique problem in an IDNC graph  (to be precisely defined in Section III). We then show that the Maximum Clique problem is equivalent to an Integer Quadratic Program (IQP) formulation. Finally, we provide a reduction from a well-known NP-Hard problem (the Exact Cover by 3-Sets) to this IQP problem.

%\footnote{\textcolor{blue}{Be careful with your words: 1-1 mapping is the same as equivalent? not the same as reduction.}}

\item We  analyze random instances of the problem, where each packet is successfully received by each user randomly and independently with the same probability. This problem, referred to as {Random Real-Time IDNC}, corresponds to a Maximum Clique problem on an appropriately created random IDNC graph.  Surprisingly, we show that when the number of packets is linear or polynomial in the number of users, the Maximum Clique problem can be solved with high probability on this particular family of random graphs, by a polynomial-time (in the number of users) algorithm, which we refer to as the {\em Max Clique} algorithm.

\squishend

We implement and compare the proposed coding scheme, Max Clique, against two baselines: an optimal repetition code and a COPE-like greedy scheme proposed in \cite{idnc-netcod08}. Simulations show that Max Clique significantly outperforms these state-of-the-art schemes over a range of scenarios, for the loss probability varying from .01 to .99. For example, for 20 users and 20 packets, Max Clique improves by a factor of 1.3 on average, and performs up to 1.6 times better than the COPE-like code and up to 3.8 times better than the optimal repetition code. Finally, we evaluate Max Clique on network traces of a real-time multi-player game on Android that uses broadcast. The results of this trace-based evaluation confirm the superior performance of Max Clique over the baselines.

The remainder of this paper is organized as follows. Section \ref{sec:related} discusses related work. Section \ref{sec:formulation} formulates the problem. Section \ref{sec:max-clique} describes the maximum clique and integer program formulations as well as the proof of NP-completeness. Section \ref{sec:random-clique} analyzes the probabilistic version (Random Real-Time IDNC problem) and describes Max Clique, the polynomial-time algorithm to find a maximum clique w.h.p. Section \ref{sec:evaluation} evaluates and compares our coding scheme with existing schemes. Section \ref{sec:conclusion} concludes the paper.

%--------------- Related Work ------------------------------------------------
\section{Related Work}
\label{sec:related}

%%% IDNC -----------------------

{\flushleft \bf  Instantly Decodable Network Coding.} Katti \ea~\cite{idnc-ton08} proposed COPE, an opportunistic inter-session network coding scheme for wireless networks. Encoded packets are chosen so that they are immediately decodable at the next hop. The algorithm considers combining packets in a FIFO way (first-in-first-out, as stored in the transmitting queue) and greedily maximizes the number of receivers that can decode in the next time slot.  Keller \ea~\cite{idnc-netcod08}
%defined decoding delay as the number of packets that are successfully received but cannot be decoded immediately. Then, they 
investigated algorithms that minimize decoding delay, including two algorithms that allow for instantaneous decoding: a COPE-like greedy algorithm and a simple repetition algorithm. In Section \ref{sec:evaluation}, we use these two algorithms as baselines for comparison. 

In \cite{idnc-netcod09}, Sadeghi \ea~improved the opportunistic algorithm previously proposed in \cite{idnc-netcod08} by giving high priority to packets that are needed by a large number of users. The authors also gave an Integer Linear Program formulation to the problem of finding an instantly decodable packet that maximizes the number of beneficiary users. 
Furthermore, they showed that it is NP-hard based on the {\em Set Packing} problem.
%They show that this formulation is equivalent to the {\em set packing} formulation, and therefore, is NP-hard. 
We note that their formulation differs from ours since it requires that a coded packet must be instantly decodable by {\em all} users, where some users may not benefit from the packet. This may lead to a suboptimal solution because there may be a coded packet that is only instantly decodable {\em by some but not all} users but is beneficial to a larger number of users. Our formulation ensures that we find this optimal packet.

Sorour \ea~have an extensive line of work investigating instantly decodable codes \cite{idnc-isit09, idnc-icc10, idnc-globecom10, idnc-icc11, idnc-pimrc11, idnc-isit12, idnc-netcod13}, focusing on minimizing the completion delay.
%{\color{blue} I think you are citing too many of their work!!!!}
They introduced the term Instantly Decodable Network Coding (IDNC) that we adopt in this work. 
In \cite{idnc-isit09}, they proposed a construction of IDNC graphs based on feedback from the users  and then introduced a transmission scheme based on graph partitioning.  We consider the same construction of IDNC graphs as in \cite{idnc-isit09}.
%Our IDNC graphs are constructed as in \cite{idnc-isit09}. 
Based on a stochastic shortest path formulation, they  proposed a heuristic algorithm to minimize the completion delay \cite{idnc-icc10}. 
In \cite{idnc-globecom10}, they introduced the notion of {\em generalized} IDNC problem, which does not require the transmitted code to be decodable by all users, as opposed to the {\em strict} version studied previously \cite{idnc-netcod08, idnc-isit09, idnc-icc10}.
Real-Time IDNC considers the generalized version.  
Furthermore, in \cite{idnc-globecom10}, they related finding an optimal IDNC code to the Maximum Clique problem in IDNC graphs and suggested that it is NP-Hard; however, no explicit reduction was provided.
%{\color{blue} So what have they shown? they have conjectured it is NP-hard or they have proven it? -- They claimed it NP-Hard without a proper proof.}. 
In \cite{idnc-icc11} and \cite{idnc-pimrc11}, they extended \cite{idnc-icc10} to cope with limited or lossy feedback. In \cite{idnc-isit12}, they considered the case of multicast instead of broadcast, and in \cite{idnc-netcod13}, the case where users could buffer coded packets in addition to plain packets was investigated.

Li \ea~\cite{idnc-jsac11} adopted IDNC for video streaming  and showed that, for independent channels and sufficiently large video file,  their proposed IDNC schemes are asymptotically throughput-optimal subject to hard deadline constraints when there are {\em no more than three users}. In contrast, we consider an arbitrary number of users, and we provide the optimal single transmission. % regardless of the subsequent transmissions.

{\flushleft \bf Index Coding.} 
Our problem setup is relatively similar to that of the Index Coding (IC) problem, introduced by Birk and Kol \cite{ic-it06} previously and extensively studied since.  An IC problem also considers a base station that knows a set of packets, $\mathcal{X}$, and a set of users. Each user $(x, \mathcal{H})$ demands one particular packet, $x \in \mathcal{X}$, and has side information consisting of a subset of packets, $\mathcal{H} \subset \mathcal{X}$. The base station broadcasts to all users without errors. The goal is to find an encoding scheme that minimizes the number of transmissions required to deliver the packets to all users. 

It has been shown that except for the cases that can be solved with one or two transmissions, other instances of the IC problem are NP-hard to solve \cite{ic-itw07, ic-globecom12, ic-isit12-maleki}, including a variation of IC where users are pliable and happy to receive any one packet \cite{picod, picod2}. Furthermore, even finding an approximation to the problem has been shown to be hard \cite{ic-isit08}. \cite{ic-infocom08, ic-isit12-arash} provided heuristic algorithms to find such codes.

Despite the similarities, there are two main differences between our problem and IC. First, in our problem, each user wants {\em all} the packets, not just a single packet. Second,  we want to find an instantly decodable packet  that maximizes the number of beneficiary users, not the total number of transmissions to satisfy all users.

{\flushleft \bf Data Exchange.} The Data Exchange (DX) problem, originally introduced by El Rouayheb \ea \cite{dx-itw10}, also has a similar setup to our problem: There is a set of packets, $\mathcal{X}$, and a set of users $\mathcal{U}$. Each user, $u \in \mathcal{U}$, knows a subset of packets, $\mathcal{H}_u \subset \mathcal{X}$, and wants all packets in $\mathcal{X}$. In DX, there is no base station, and the users broadcast messages.  The objective is to find an encoding scheme that minimizes the number of transmissions required to deliver all packets in $\mathcal{X}$ to all users.

To solve the DX problem, a randomized polynomial-time solution was proposed in \cite{dx-isit10}, and deterministic polynominal-time solutions were proposed in \cite{dx-qshine10} and \cite{dx-isit11}. \cite{dx-isit12-g} studied the problem in general network topologies. Variants of the problem where there are helpers and transmission weights were studied in \cite{dx-isit12} and \cite{dx-ita11}. Various necessary and sufficient conditions that characterize feasible transmission schemes for the problem were proposed in \cite{uc-milcom10, uc-allerton10, uc-allerton11}, under a different name of {\em universal recovery}.

Similar to DX, in our setting, all users want all the packets in $\mathcal{X}$. However, there are two main differences: (i) in our  setting, only the base station can broadcast as opposed to having all users capable of broadcasting, and (ii) we are interested in instantaneous decoding to maximize the number of beneficiary users with one transmission, as opposed to minimizing the total number of transmissions.

%We highlight that in this work, we show that the Real-Time IDNC problem is NP-Hard, similar to the IC problem. The DX problem, however, has polynomial-time solutions.

{\flushleft \bf This Work in Perspective.} A preliminary version of this work has appeared in \cite{le-netcod13}. In this paper, we extend the previous work in the following ways: First, we provide complete proofs of all theorems and propositions. Second, we collect network traces of a real-time application that utilizes broadcast and present a new evaluation based on the traces. Finally, we discuss and highlight the similarities and differences between the Real-Time IDNC, Index Coding, and Data Exchange problems.

%--------------- Problem Formulation --------------------------------------------------
\section{Problem Formulation}
\label{sec:formulation}

Let $\uset = \{u_1, \cdots, u_n\}$ denote the set of $n$ users, and $\pset = \{p_1, \cdots, p_m\}$ be the set of $m$ packets. We assume that the original $m$ packets were broadcast by a base station. Due to packet loss, each of $n$ users missed some of the $m$ packets. We denote the set of packets that were successfully received by user $i$ by $\hset{i}$. Furthermore, let $\wset{i}$ be the set of packets that user $i$ still wants, \ie, $\wset{i} = \pset \setminus \hset{i}$. Consistently with 
\cite{ic-itw07, idnc-isit09}, we call $\mathcal{H}$'s and $\mathcal{W}$'s the ``Has'' and ``Want'' sets.

After the initial broadcast, the base station tries to recover the losses, $\mathcal{W}$'s, by sending coded packets and exploiting the side information  of the already delivered packets, $\mathcal{H}$'s.
Let the $n \times m$ matrix $\matr{A}$ be the identification matrix for the side information of the users, \ie, entry $a_{ij} = 1$ if user $u_i$ wants packet $p_j$ and 0 otherwise. $\matr{A}$ is also called a feedback matrix, as in \cite{idnc-isit09, idnc-icc10, idnc-globecom10, idnc-icc11, idnc-pimrc11, idnc-isit12}. Let us clarify this by an example. 

\begin{figure}[tp]
\centering
\includegraphics[width=7cm]{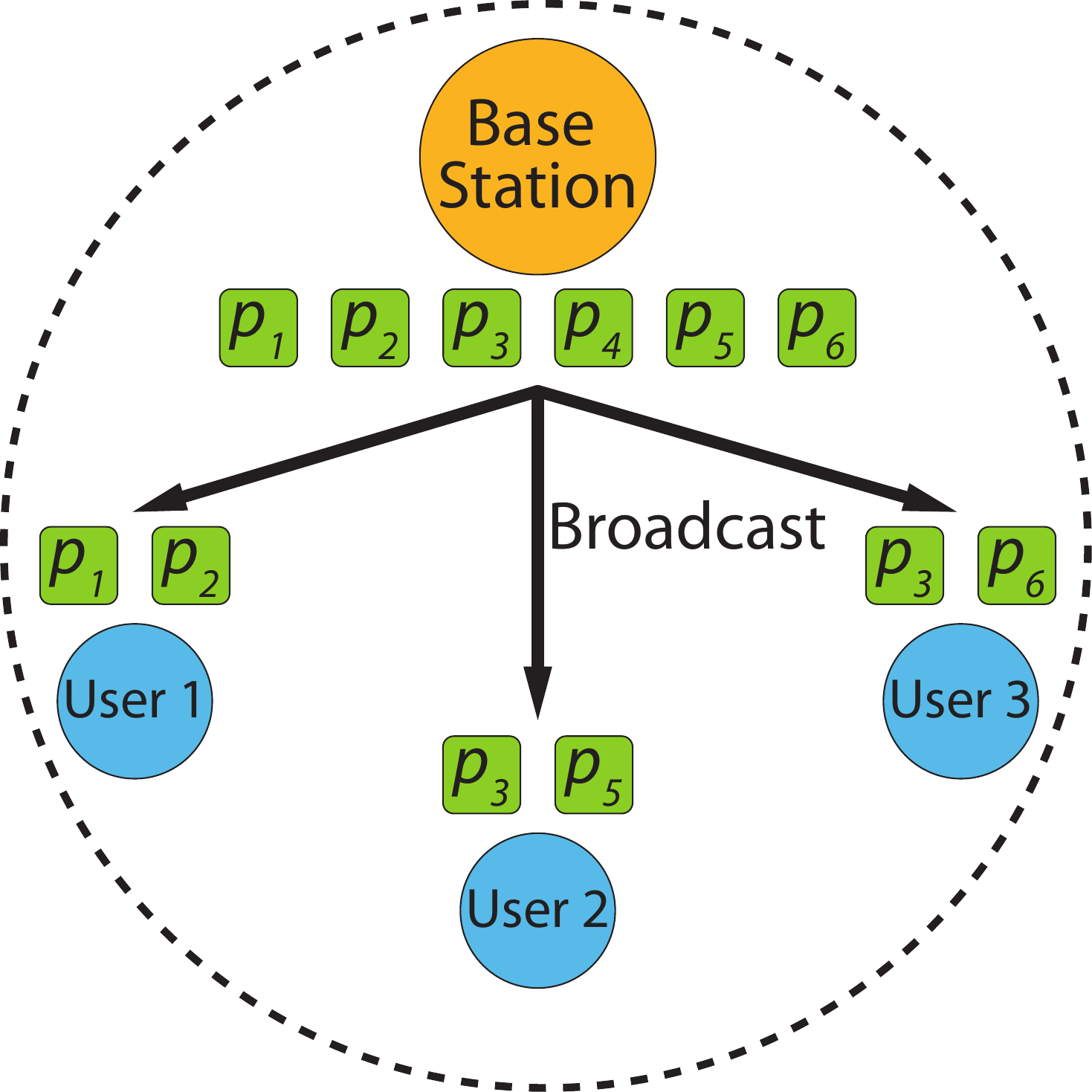}
\vspace*{5pt}
\caption{Example 1: A base station broadcast 6 packets $\{p_1, \cdots, p_6\}$ to 3 users. Due to packet loss, user 1 only received $p_1$ and $p_2$; user 2 received $p_3$ and $p_5$; user 3 received $p_3$ and $p_6$.\label{fig:broadcast}}
\end{figure}

\begin{example}
Consider a scenario with 3 users and 6 packets. Furthermore, assume that after the initial broadcast, user $u_1$ successfully received packets $p_1$ and $p_2$; user $u_2$ received $p_3$ and $p_5$; and user $u_3$ received $p_3$ and $p_6$. The scenario is depicted in Fig. \ref{fig:broadcast}. In this case, the side information matrix is as follows:
\[
\matr{A} = \left(
\begin{array}{cccccc}
0 & 0 & 1 & 1 & 1 & 1\\
1 & 1 & 0 & 1 & 0 & 1\\
1 & 1 & 0 & 1 & 1 & 0
\end{array}
\right)\,.
\]
\label{ex1}
\end{example}

To deliver the packets in the Want sets of the users, we focus on instantly decodable, lightweight coding schemes that operate in GF(2).  For a set of packets, $\mathcal{M}$, the corresponding coded packet $c$ is their binary sum, denoted by $\bigoplus$: 
\begin{eqnarray}\label{codedpacket}
c = \bigoplus_{p_i \in \mathcal{M}}p_i\,.
\end{eqnarray}
%where $\bigoplus$ denotes the binary sum.

%\footnote{ \textcolor{blue}{The term ``benefit'' was not defined properly anywhere. Initially I thought it is the same as "innovative", but they are different.}}

\begin{definition}
A coded packet, $c^{\mathcal{N}}$, is {\em instantly decodable} with respect to a set of users, $\mathcal{N}$, if and only if
\begin{enumerate}[(i)]
\item Every user, $u_i \in \mathcal{N}$, can decode $c^{\mathcal{N}}$ immediately upon reception to recover a packet $p^i \in \wset{i}$. That is, each user in $\mathcal{N}$ {\em benefits} from $c^{\mathcal{N}}$ by recovering one of the packets from its want set.
\item Every packet in the binary sum of $c^{\mathcal{N}}$ is wanted by at least one user in $\mathcal{N}$. 
%Each packet that is a component of $c^{\mathcal{N}}$ must be wanted by at least one user in $\mathcal{N}$.
\end{enumerate}
\end{definition}

%For example, let $\oplus$ denote the binary XOR operation. 
For example, for the scenario of Example \ref{ex1}, the coded packet $c^{\{u_1, u_2, u_3\}} = p_1 \oplus p_3$ is instantly decodable with respect to $\{u_1, u_2, u_3\}$ since $u_1$ can recover $p_3$, while $u_2$ and $u_3$ can get $p_1$. Meanwhile, $c^{\{u_2, u_3\}} = p_5 \oplus p_6$ is not instantly decodable with respect to $u_1$. Furthermore, we do not consider $c^{\{u_2, u_3\}} = p_1 \oplus p_5 \oplus p_6$ instantly decodable with respect to $\{u_2, u_3\}$ since although $c$ can be decoded by $u_2$ and $u_3$, packet $p_1$, which is a component of $c_3$, is not needed by either $u_2$ or $u_3$. (From here on, we will omit the superscript $\mathcal{N}$ of $c^\mathcal{N}$ when there is no ambiguity.)

We would like the coded packet to be  immediately beneficial to as many users as possible. Thus, our notion of optimality is with respect to the cardinality of the set of beneficiary users  $|\mathcal{N}|$.
%\begin{definition}
%We call an instantly decodable code $c^{\mathcal{N}}$ {\em optimal} if and only if there does not exist an instantly decodable code $c^{\mathcal{N}'}$ for which $|\mathcal{N}'| > |\mathcal{N}|$.  \footnote{\textcolor{blue}{Do you need a separate definition or could you merge with either  problem statement below or text above?}}
%\end{definition}
{\flushleft \bf The Real-Time IDNC Problem:} Given a side information matrix $\matr{A}$, find  the optimal instantly decodable packet $c^\mathcal{N}$.

%---------------  Maximum Clique in IDNC --------------------------------------------------
\section{Maximum Cliques in IDNC Graphs}
\label{sec:max-clique}

%{\bf \textcolor{blue}{[Athina: I commented out graph.tex due to compile errors. Remember to put it back in.]}}
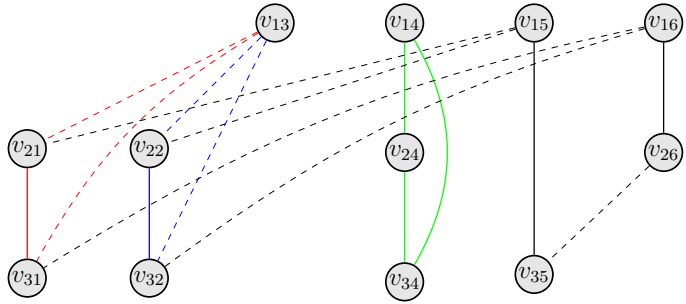
\begin{figure}[tp]
\centering
\begin{tikzpicture}
\node[scale=0.8] {% scale the picture here
\begin{tikzpicture}

\tikzstyle{solid} =[
	semithick]

\tikzstyle{dash}=[
	dashed]

\tikzstyle{circ} =[
	circle,
	draw=black,
	fill=black!10,
	thick,
	minimum width = 5mm,
	inner sep=0.2mm]

\tikzstyle{invicirc} =[
	circle,
	draw=white,
	thick,
	minimum width = 5mm,
	inner sep=0.2mm]

% Set of vertices
\node[invicirc] (v11) at (0,0) {};
\node[invicirc, right=15mm of v11] (v12) {};
\node[circ, right=15mm of v12] (v13) {$v_{13}$};
\node[circ, right=15mm of v13] (v14) {$v_{14}$};
\node[circ, right=15mm of v14] (v15) {$v_{15}$};
\node[circ, right=15mm of v15] (v16) {$v_{16}$};

\node[circ, below=15mm of v11] (v21) {$v_{21}$};
\node[circ, below=15mm of v12] (v22) {$v_{22}$};
\node[invicirc, below=15mm of v13] (v23) {};
\node[circ, below=15mm of v14] (v24) {$v_{24}$};
\node[invicirc, below=15mm of v15] (v25) {};
\node[circ, below=15mm of v16] (v26) {$v_{26}$};

\node[circ, below=15mm of v21] (v31) {$v_{31}$};
\node[circ, below=15mm of v22] (v32) {$v_{32}$};
\node[invicirc, below=15mm of v23] (v33) {};
\node[circ, below=15mm of v24] (v34) {$v_{34}$};
\node[circ, below=15mm of v25] (v35) {$v_{35}$};
\node[invicirc, below=15mm of v26] (v36) {};

% Set of solid edges
\draw[solid, draw=red] (v21) to (v31);
\draw[solid, draw=blue] (v22) to (v32);
\draw[solid, draw=green] (v14) to (v24);
\draw[solid, draw=green] (v24) to (v34);
\draw[solid, draw=green, bend left] (v14) to (v34);
\draw[solid] (v15) to (v35);
\draw[solid] (v16) to (v26);

% Set of dashed edges
\draw[dash, draw=red] (v13) to (v21);
\draw[dash, draw=blue] (v13) to (v22);
\draw[dash, bend right=15, draw=red] (v13) to (v31);
\draw[dash, draw=blue] (v13) to (v32);
\draw[dash] (v15) to (v21);
\draw[dash] (v15) to (v22);
\draw[dash, bend right=10] (v16) to (v31);
\draw[dash, bend right=10] (v16) to (v32);
\draw[dash] (v26) to (v35);

\end{tikzpicture}
};
\end{tikzpicture}
\vspace*{5pt}
\caption{The Instantly Decodable Network Coding (IDNC) graph of Example \ref{ex1}. Solid edges are edges of type (i) and dashed edges are edges of type (ii). There are three maximum cliques: $\{v_{13}, v_{21}, v_{31}\}$, $\{v_{13}, v_{22}, v_{32}\}$, and $\{v_{14}, v_{24}, v_{34}\}$, all of which are of size 3.}
\label{fig:graph}
\end{figure}

%--------------- IDNC Graph --------------------------------------------------
Given a side information matrix $\matr{A}$, we form an Instantly Decodable Network Coding (IDNC) graph corresponding to $\matr{A}$ as in \cite{idnc-isit09}: We create a vertex $v_{ij}$ when user $u_i$ still wants packet $p_j$. For instance, for matrix $\matr{A}$ in Example \ref{ex1}, there is a vertex for each entry 1 in the matrix. Given a vertex $v_{ij}$, we use the term {\em user index} of $v_{ij}$ to indicate $i$ and {\em packet index} of $v_{ij}$ to indicate $j$. There is an edge between two vertices $v_{ij}$ and $v_{k\ell}$ if one of the below conditions hold:
\begin{enumerate}[(i)]
\item $j = \ell$: In this case, both users $u_i$ and $u_k$ wants the same packet $p = p_j = p_\ell$.
\item $p_j \in \hset{k}$ and $p_{\ell} \in \hset{i}$: In this case, user $u_k$ has packet $p_j$ that user $u_i$ still wants, and vice versa.
\end{enumerate}
Denote the IDNC graph corresponding to a matrix $\matr{A}$ by $G^\matr{A} = (\mathcal{V}, \mathcal{E})$. Figure \ref{fig:graph} shows the IDNC graph corresponding to the side information matrix given in Example \ref{ex1}. 

%--------------- Mapping --------------------------------------------------
\subsection{Cliques and Instantly Decodable Packets}

\begin{proposition}\label{proposition:optimal-code-max-clique}
Finding an optimal instantly decodable code given a side information matrix $\matr{A}$ is equivalent to finding a maximum clique in the corresponding IDNC graph $G^\matr{A}$.
\end{proposition}

We prove this proposition by establishing the following Lemmas \ref{lemma:clique-to-code} and \ref{lemma:number-user}. The first lemma states the relationship between instantly decodable packets and cliques in $G^\matr{A}$.

%------- Start: lemma clique-to-code -------
\begin{lemma}\label{lemma:clique-to-code}
Given a side information matrix $\matr{A}$ and its IDNC graph $G^A$, an instantly decodable packet has a one-to-one correspondence to a clique in $G^A$.
\end{lemma}
%------- End: lemma clique-to-code -------
%\end{comment}

The second lemma expresses the relationship between the number of users benefiting from an instantly decodable packet and the size of the clique corresponding to the packet.

%------- Start: lemma number user -------
\begin{lemma}\label{lemma:number-user}
Given a side information matrix $\matr{A}$ and its IDNC graph $G^\matr{A}$, let $c^\mathcal{N}$ be an instantly decodable packet, and let $\mathcal{C}$ be the corresponding clique of $c^\mathcal{N}$ in $G^A$, then $|\mathcal{C}| = |\mathcal{N}|$.
\end{lemma}
%------- End: lemma number user -------

\begin{comment}
The below corollary follows from Lemmas \ref{lemma:clique-to-code} and \ref{lemma:number-user}.
\begin{corollary}\label{corollary:code-to-clique}
Given a side information matrix $A$ and its IDNC graph $G^A$, an optimal instantly decodable packet $c^\mathcal{N}$ has a one-to-one correspondence to a maximum clique $\mathcal{C}$ in $G^A$, and $|\mathcal{C}| = |\mathcal{N}|$.
\end{corollary}
\end{comment}

The proofs of Lemma \ref{lemma:clique-to-code} and \ref{lemma:number-user} are provided in Appendices \ref{appendix:lemma2} and \ref{appendix:lemma3}, correspondingly. Intuitively, let us consider the clique involving $v_{13}$, $v_{21}$, and $v_{31}$ in Example 1. XORing all packets corresponding to vertices of this clique, \ie, $p_1 \oplus p_3$, forms an instantly decodable packet because (i) user 1 must have $p_1$, and users 2 and 3 must have $p_3$, otherwise there are no edges ($v_{13}$, $v_{21}$) and ($v_{13}$, $v_{31}$), and (ii) each component of the coded packet is wanted by the user corresponding to the row of the vertex. Finally, the clique size equals 3, which is the number of beneficiary users.

%--------------- NP-Completeness --------------------------------------------------
\subsection{NP-Completeness}
Finding a maximum clique in a general graph is well known to be NP-Hard. This result, however, is not directly applicable to IDNC graphs as they have special structural properties. In this section, we will show that the problem of finding a maximum clique in an IDNC graph is indeed NP-Hard. We show this by first showing that finding a maximum clique in an IDNC graph is equivalent to finding an optimal solution to an Integer Quadratic Programming (IQP) problem. We then describe a reduction from a well known NP-Complete problem, the Exact Cover by 3-Sets problem, to the decision version of the IQP problem.

%--------------- Quadratic Programming Formulation ----------------------
\subsubsection{Integer Quadratic Programming Formulation}
Given a side information matrix $\matr{A}$ of size $n \times m$, we formulate the IQP problem as follows. Let $\vct{r}$ be a binary $n \times 1$ vector: $r_i \in \{0,1\}, i = 1, \cdots, n$. Similarly, let $\vct{c}$ be a binary $m \times 1$ vector: $c_j \in \{0,1\}, j = 1, \cdots, m$. Below is the IQP problem for $\matr{A}$:
\begin{center}
\begin{tabular}{|lll|}
\hline
~&~&~\\
Maximize: & $V = \vct{r}^T \, \matr{A} \, \vct{c} = \sum_{i=1}^n \, \sum_{j=1}^m r_i \, c_j \, a_{ij}  $.&~\\
~&~&~\\
Subject to: & $r_i \, \sum_{j=1}^m c_j \, a_{ij} \leq 1, \, \forall i = 1, \cdots, n\,.\quad$ & (1)\\
~ & $ r_i, c_j \in \{0, 1\}\,.\quad$ & (2)\\
~&~&~\\
\hline
\end{tabular}
\vspace*{20pt}
\end{center}

\begin{proposition}\label{proposition:max-clique-optimal-iqp}
Given a side information matrix $\matr{A}$ and its IDNC graph $G^A$, finding a maximum clique in $G^\matr{A}$ is equivalent to finding an optimal solution to the corresponding IQP.
\end{proposition}
We prove this proposition by establishing the following Lemmas \ref{lemma:clique-to-iqp} and \ref{lemma:iqp-value}. The first lemma expresses the relationship between the above IQP problem and the problem of finding maximum clique in $G^\matr{A}$.

\begin{lemma}\label{lemma:clique-to-iqp}
Given a side information matrix $\matr{A}$ and its IDNC graph $G^\matr{A}$, a clique in $G^\matr{A}$ has a one-to-one correspondence to a feasible solution of the IQP problem for $\matr{A}$.
\end{lemma}
\begin{IEEEproof}
{\flushleft ($\Rightarrow$)} We first show that a clique in the IDNC graph maps to a feasible pair of vectors $\vct{r}$ and $\vct{c}$ of the IQP problem, which is uniquely identified by the user and packet indices of the vertices in the clique.

Let $\mathcal{C}$ be a clique in $G^\matr{A}: \mathcal{C} = \{v_{i_1,j_1}, \cdots, v_{i_k,j_k}\}$. Let $\mathcal{I}$ be the set of user indices: $\mathcal{I} = \{i_1, \cdots, i_k\}$, and $\mathcal{J}$ be the set of packet indices: $\mathcal{J} = \{j_1, \cdots, j_k\}$. We create the feasible pair of $\vct{r}$ and $\vct{c}$ as follows: Set $r_i = 1$ if $i \in \mathcal{I}$ and 0 otherwise, and set $c_j = 1$ if $j \in \mathcal{J}$ and 0 otherwise. 

To show that this pair of vectors is a feasible solution, we proceed by showing that condition (1) of the IQP holds for all user indices. Let $i$ be any user index, $i \in [1,n]$. It is clear that (1) holds if $r_i = 0$. When $r_i = 1$, it suffices to show that no two vertices of $\mathcal{C}$ have the same user index. Indeed, this follows from the observation that there is no edge between any two vertices having the same user index (on the same row) in the IDNC graph. 

{\flushleft ($\Leftarrow$)} Next, we show that a feasible solution of the IQP  maps to a uniquely identified clique in the IDNC graph. Let the pair of vectors $\vct{r}$ and $\vct{c}$ be a feasible solution. We map this pair to a clique $\mathcal{C}$ in $G^\matr{A}$ as follows: Initialize $\mathcal{C} = \emptyset$. For $i \in [1,n]$, for $j \in [1,m]$, if $r_i = c_j = a_{ij}=1$, add vertex $v_{ij}$ to $\mathcal{C}$.

Now pick any pair of vertices $v_{st}$ and $v_{pq}$ in $\mathcal{C}$. It is clear that if $t=q$, there is an edge between these two vertices. It remains to show that when $t \neq q$, user $u_s$ has  packet $p_q$ and user $u_p$ has packet $p_t$. We will show that user $u_s$ must have  packet $p_q$. The other condition follows by symmetry. Assume otherwise, \ie, user $u_s$ does not have packet $p_q$, which means $a_{sq} = 1$. Since $\mathcal{C}$ contains $v_{st}$ and $v_{pq}$, $r_s = c_t = a_{st}=1$ and $r_p = c_q = a_{pq}=1$. But then for row $s$, condition (1) of the IQP problem fails since
\[ r_s \, \sum_{j=1}^m c_j \, a_{sj} \geq  r_s \, c_t \, a_{st} + r_s \, c_q \, a_{sq} = 2\,.\]

Finally, it is easy to check that for the above two mappings, one is the reverse of the other.
\end{IEEEproof}

%It turns out that the size of a clique in $G^A$ also has a strong relationship with the value $V$ of the objective function of the IQP problem, as stated in the below lemma.

\begin{lemma}\label{lemma:iqp-value}
Given a side information matrix $\matr{A}$ and its IDNC graph $G^\matr{A}$, the size of a clique in $G^\matr{A}$ equals to the objective value $V$ of its corresponding feasible solution of the IQP problem for $\matr{A}$.
\end{lemma}
\begin{IEEEproof}
Let $\mathcal{C}$ be a clique in $G^\matr{A}$ and $\vct{r}$ and $\vct{c}$ be the pair of vector of the corresponding feasible solution. For any user index $i$ and packet index $j$, if $v_{ij} \in \mathcal{C}$, then $r_i = c_j = a_{ij} = 1$. Hence, every vertex in the clique adds 1 to $V$. 
\end{IEEEproof}

%--------------- Reduction from Exact Cover by 3-Sets ----------------------
\subsubsection{Reduction from Exact Cover by 3-Sets}
Given a side information matrix $A$, the decision version of the IQP problem for $\matr{A}$, denoted as {\em D-IQP}, asks the following question: ``Is there a feasible solution whose objective value equals $N$, for some $N >0$?'' 

\begin{proposition}\label{proposition:diqp-np-complete}
The D-IQP problem is NP-Complete.
\end{proposition}

\begin{IEEEproof}
Clearly, D-IQP is in NP since given a feasible pair of vectors $\vct{r}$ and $\vct{c}$, we can compute the objective value in polynomial $O(nm)$ time. 

In what follow, we show a reduction from the Exact Cover by 3-Sets (X3C) problem to D-IQP. X3C is well-known to be an NP-Complete problem \cite{np-hard} and is defined as follows:

\begin{definition}
Given a set $\mathcal{E}$ of $3 k$ elements: $\mathcal{E} = \{e_1, \cdots, e_{3k}\}$, and a collection $\mathcal{F} = \{ \mathcal{S}_1, \cdots, \mathcal{S}_\ell \} $ of subsets $\mathcal{S}_i \subset \mathcal{E}$ and $|\mathcal{S}_i| = 3$, for $i \in [1,\ell], \ell > k$. The X3C problem asks the following question: ``Are there $k$ sets in $\mathcal{F}$ whose union is $\mathcal{E}$?''
\end{definition}

{\flushleft \bf The reduction:} Given any instance of X3C, we create $3k$ users, $u_1, \cdots, u_{3k}$, and $\ell$ packets, $p_1, \cdots, p_\ell$. The users correspond to the elements $e_i, i \in [1,3k],$ and the packets correspond to the sets $S_j, j \in [1,\ell]$. We form the side information matrix $\matr{A}^\text{X3C}$ corresponding to this X3C instance by setting $a_{ij} = 1$ if $e_i \in \mathcal{S}_j$ and 0 otherwise.

Next, we will show that there is a feasible solution to the D-IQP for $\matr{A}^\text{X3C}$ whose objective value $V$ equals $3k$ if and only if there are $k$ sets $\mathcal{S}_{j_1}, \cdots, \mathcal{S}_{j_k}$ whose union is $\mathcal{E}$.

{\flushleft ($\Rightarrow$)} Let $\vct{r}$ and $\vct{c}$ be the pair of vectors of the feasible solution whose objective value $V = 3k$. First, observe that all $r_i$, for $i = 1, \cdots, 3k,$ must equal 1; otherwise, assume there exists some index $t \in [1, 3k]$ where $r_t = 0$, then 
\[ V = \sum_{i=0}^{3k} r_i \sum_{j=0}^\ell c_j a_{ij} = \sum_{i=0, i \neq t}^{3k} r_i \sum_{j=0}^\ell c_j a_{ij} < 3k\,,\]
since each term $r_i \sum_{j=0}^\ell c_j a_{ij}$ is at most 1 by constraint (1). This is a contradiction.

Next, we create the corresponding solution to the X3C problem using $\vct{c}$. In particular, for $j=1, \cdots, \ell$, we select $S_j$ if $c_j = 1$. Because $V = 3k$ and $r_i =1$ for all $i$, it must be that
\[ \sum_{j=1}^\ell c_j a_{ij} = 1,\quad  \text{ for } i = 1, \cdots, 3k. \]
Thus, for a user index $s \in [1,3k]$, there exists a {\em unique} packet index $t \in [1,\ell]$, where $c_t \, a_{st} = 1$, which means $c_t = a_{st}=1$. By construction, we selected set $S_t$, and this $S_t$ covers element $s$ as  $a_{st} = 1$ . Therefore, every element is contained in exactly one set.

{\flushleft ($\Leftarrow$)} Let $\mathcal{S}_{j_1}, \cdots, \mathcal{S}_{j_k}$ be the solution to the X3C problem. We create the corresponding solution to the D-IQP problem as follows. First, for $\vct{r}$, let $r_i = 1$, for all $i = 1, \cdots, 3k$. Then, for $\vct{c}$, let $\mathcal{J} = \{j_1, \cdots, j_k\}$, and for $j = 1, \cdots, \ell$, set $c_j = 1$ if $j \in \mathcal{J}$ and 0 otherwise. Since $\mathcal{S}_{j_1}, \cdots, \mathcal{S}_{j_k}$ covers all $3k$ elements and each set has only 3 elements, each element $e_s$ appears in exactly one set $\mathcal{S}_{j_t}$ for some $t \in [1,k]$, and $c_{j_t} = 1$. Thus, for each element $s \in [1,3k]$,
\[ \sum_{j=0}^\ell c_j a_{sj} = c_{j_t} a_{s\, {j_t}} = 1 \cdot 1 = 1 \]
Given the above $\vct{r}$ and $\vct{s}$,
\[ V = \sum_{i=0}^{3k} r_i \sum_{j=0}^\ell c_j a_{ij} = 3k \cdot 1 = 3k\,. \]

\end{IEEEproof}

From Propositions \ref{proposition:optimal-code-max-clique}, \ref{proposition:max-clique-optimal-iqp}, and \ref{proposition:diqp-np-complete}, we have the following main result of this work.

\begin{theorem}
Given a side information matrix $\matr{A}$ and its IDNC graph $G^\matr{A}$, finding a maximum clique in $G^\matr{A}$, and equivalently, an optimal instantly decodable packet, is NP-Hard. Their corresponding decision versions are NP-Complete.
\end{theorem}

%--------------- Random IDNC --------------------------------------------------
\section{Maximum Cliques in Random IDNC Graphs}
\label{sec:random-clique}

In this section, we investigate Random Real-Time IDNC. In particular, we assume that each user, $u_i$,  $i \in [1,n]$, fails to receive a packet, $p_j$, $j \in [1,m]$, with the same probability, $p \in (0,1)$, independently. For ease of analysis, we assume that $m$ is linear in $n$: $m= d\,n$, for some constant $d > 0$. (Our results also hold when $m$ is polynomial in $n$.) 

A random IDNC graph, denoted as $G^\matr{A}(p)$, is the graph corresponding to a side information matrix $\matr{A}$, whose each entry equals 1 with probability $p$ and 0 with probability $q = 1-p$ independently. Next, we will provide the analysis of the size of the maximum clique, \ie, the clique number, of random IDNC graphs. 

The main results of this section are the followings:
\begin{enumerate}[(i)]
\item For any $p \in (0,1)$, the clique number for almost every graph in $G^\matr{A}(p)$ is linear in $n$. In particular, it equals $j^* p q^{j^* -1} n$, where $j^* = \text{argmax } j p q^{j-1}$, $j^* \in \mathbb{N}$. With high probability, the optimal recovery packet involves combining $j^*$ packets. 
%An important implication is that for small $p$ $(p < 0.5)$, combining of at least $2$ packets is necessary to achieve the optimal encoding.

\item With high probability, the maximum clique can be found in polynomial time, $O (n\,m^{j^* + \delta})$, where $\delta$ is a small constant parameter,  and we provide an explicit algorithm, Max Clique, to find it. Consequently, the optimal recovery packet can be computed in polynomial time in $n$.
\end{enumerate}

{\flushleft \bf Comparison to  Erd\H{o}s-R\'enyi Random Graphs:} Clique numbers of Erd\H{o}s-R\'enyi random graphs with $n$ vertices and $p=1/2$ are known to be close to $2 \log_2 n$ \cite{clique-text}. However, it is widely conjectured that for any constant $\epsilon > 0$, there does not exist a polynomial-time algorithm for finding cliques of size $(1+\epsilon) \log_2 n$ with significant probability \cite{clique-soda98}. In contrast, for random IDNC graphs with $n \times m$ vertices, where $m$ is linear or polynomial in $n$, we show  that the clique numbers are linear in $n$, and the corresponding cliques can be found in polynomial time in $n$.

%------------------------------------------
\subsection{Clique Number of Random IDNC Graphs}
\label{subsec:clique-number-random-idnc}

First, observe that any $k$ 1's that lie in the same column form a clique of size $k$. Since the expected number of 1's in a single column is $n p$, the expected size of single-column cliques is $n p$. As a result, we expect the maximum clique size to be linear in $n$.

Fix a set $\mathcal{C}_j$ of $j$ columns. A row $r$ is said to be {\em good} with respect to $\mathcal{C}_j$ if 
among the $j$ columns, it has 1 one and $j-1$ zeros. The probability that a row is good w.r.t. $\mathcal{C}_j$ is 
\begin{align}
 f(j) = j \, p \, q ^ {j-1}\,. \label{eqn:pgood}
\end{align}
Let $Z_{\mathcal{C}_j}$ be the number of good rows w.r.t $\mathcal{C}_j$. Then $Z_{\mathcal{C}_j}$ has a binomial distribution: $\text{Bin}(n, f(j))$.

Let $X_{\mathcal{C}_j}$ be the size of the maximum clique that has at least one vertex on every column in $\mathcal{C}_j$, \ie, the clique touches $j$ columns. Observe that if $j=1$, then $f(1) = p$, and $X_{\mathcal{C}_1} = Z_{\mathcal{C}_1}$, which is the number of 1's in the chosen column. Thus, $X_{\mathcal{C}_1}$ has a Binomial distribution: $\text{Bin}(n, p)$. For $j > 1$, $X_{\mathcal{C}_j} \neq Z_{\mathcal{C}_j}$ since the set of good rows may not have a 1 in every column in $\mathcal{C}_j$. The following lemma states that for a large $k$, where $k \overset{def}{\text{=}} Z_{\mathcal{C}_j}$ $\sim \text{Bin}(n,f(j))$, \ie, given large enough $n$, $X_{\mathcal{C}_j} = Z_{\mathcal{C}_j}$ with high probability.

\begin{lemma}\label{lemma:XZ}
For a set of constant $j$ columns $\mathcal{C}_j$, there exists a constant $k_j > 0$ such that for all $k \geq k_j$, 
\[ \text{Pr}[Z_{\mathcal{C}_j} = X_{\mathcal{C}_j} \, | \, Z_{\mathcal{C}_j} = k ] \geq 1 - j \, \left(\frac{j-1}{j}\right)^{k}\,.\]
\end{lemma}

\begin{IEEEproof}
For $k \geq j >0$, let $B^j_k$ denote the number of ways to put $k$ 1's into a matrix of size $k \times j$ such that (i) each row has one 1, and (ii) each column  has at least one 1. Note that $B^1_k = 1$, and we have the following recurrence:
\begin{align}
B^j_k &= j^k - {j \choose 1} B^{j-1}_k - {j \choose 2} B^{j-2}_k \cdots - {j \choose j-1} B^{1}_k \label{recurrence} \,.
\end{align}

This recurrence states that the number of ways to put $k$ 1's into $k$ rows (each row has one 1) using exactly $j$ columns equals to the number of ways to put $k$ 1's into $k$ rows without any column restriction subtracts the cases where there are $1, 2, \cdots,j-1$ empty columns. It can be shown by induction (details are in Appendix \ref{appendix:lemma9}) that 

\[B^j_k = \sum_{i=0}^{j-1} (-1)^i \, {j \choose i} (j-i)^k \,.\] 

Thus, we have that 

\[B^j_k = j^k - j(j-1)^k + \sum_{i=2}^{j-1} (-1)^i \, {j \choose i} (j-i)^k \,.\]

Let $k_j$ be the minimum positive integer value of $k$ such that $\sum_{i=2}^{j-1} (-1)^i \, {j \choose i} (j-i)^k \geq 0$. %For simple notation, assume this sum equals 0 when $j = 2$. 
Then, for all $k \geq k_j$, 

\[ \text{Pr}[Z_{\mathcal{C}_j} = X_{\mathcal{C}_j} \, | \, Z_{\mathcal{C}_j} = k ] = \frac{B_k^j}{j^k} \geq \frac{j^k - j(j-1)^k}{j^k}\,.\]
\end{IEEEproof}

The following lemma states that $X_{\mathcal{C}_j}$, the size of the maximum clique that touches all $j$ columns, heavily concentrates around $n f(j)$ for large $n$.

\begin{lemma}\label{lemma:boundXCj}
For a set of constant $j$ columns $\mathcal{C}_j$ and any constant $c > 1$, let $\mu = n f(j)$ and $\delta = \sqrt{\frac{3 c \, \ln n}{n \, f(j)}}$. 
%($\mu \delta = \sqrt{ 3 c f(j) \, n \ln n }$.)  = O( squareroot n logn)
For a large $n$ such that  $\mu - \mu \delta \geq k_j$ ($k_j$ is as in Lemma \ref{lemma:XZ}), we have
\[ \text{Pr}[ \,|\, X_{\mathcal{C}_j} - \mu  \,|\, \geq \mu \delta ] \leq \frac{2}{n^c} + 2\mu \delta \,  j(1-\frac{1}{j})^{\mu - \mu \delta} \,. \]
This probability goes to 0 as $n \rightarrow \infty$.
\end{lemma}

The proof is provided in Appendix \ref{appendix:lemma10}. Intuitively, this result follows from $X_{\mathcal{C}_j} = Z_{\mathcal{C}_j}$ w.h.p. (Lemma \ref{lemma:XZ}), and the fact that the Binomial distributed $Z_{\mathcal{C}_j}$, the number of good rows, concentrates heavily around its mean,  $n f(j)$. Note that $\mu \delta$ is $\Theta(\sqrt{n \ln n})$; thus, $X_{\mathcal{C}_j}$ is within  $\Theta(\sqrt{n \ln n})$ of $n f(j)$ w.h.p. 

Next, for a constant $j$, let $X_j$ be the size of the maximum clique that touches {\em any} $j$ columns. $X_j$ also heavily concentrates around $n f(j)$. Recall that $m = dn$, for some constant $d>0$. Formally,

\begin{theorem}\label{thm:boundXj}
For a constant $j$ and any constant $c > j$, let $\mu = n f(j)$ and $\delta = \sqrt{\frac{3 c \, \ln n}{n \, f(j)}}$. 
%Thus, $\mu \delta = \sqrt{ 3 c f(j) \, n \ln n }$ = O( squareroot n logn)
For a large $n$ such that  $\mu - \mu \delta \geq k_j$ ($ k_j$ is as in Lemma \ref{lemma:XZ}), we have
\[ \text{Pr}[ \,|\, X_j - \mu  \,|\, \geq \mu \delta ] \leq \frac{2 d^j }{n^{c-j}} + 2 d^j n^j \mu \delta \,  j(1-\frac{1}{j})^{\mu - \mu \delta} \,. \]
This probability goes to 0 as $n \rightarrow \infty$.
\end{theorem}

\begin{IEEEproof} The proof is by using the union bound on the result of Lemma \ref{lemma:boundXCj}:
\begin{align*}
\text{Pr}[ \,|\, X_j - \mu  \,|\, \geq \mu \delta ] &= \text{Pr}[ \cup_ {\mathcal{C}_j}  \,|\, X_{\mathcal{C}_j} - \mu  \,|\, \geq \mu \delta ] \\
~&\leq {m \choose j} \text{Pr}[ \,|\, X_{\mathcal{C}_j} - \mu  \,|\, \geq \mu \delta ] \\
~&\leq m^j \left(\frac{2}{n^c} + 2\mu \delta \,  j(1-1/j)^{\mu - \mu \delta} \right) \\
~&\leq \frac{2 d^j }{n^{c-j}} + 2 d^j n^j \mu \delta \,  j(1-1/j)^{\mu - \mu \delta}\,.
\end{align*}

\end{IEEEproof}

\begin{figure}[tp]
\centering
\includegraphics[width=8.5cm]{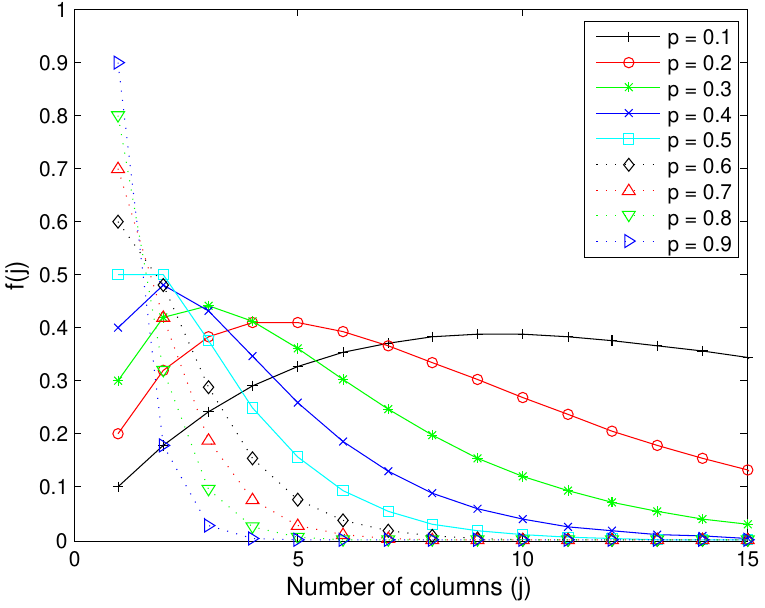}
\caption{Plot of $f(j) = j p (1-p)^{j-1}$ for different loss rate $p$\label{fig:fj}}
\vspace*{10pt}
\end{figure}

% Clique size fig
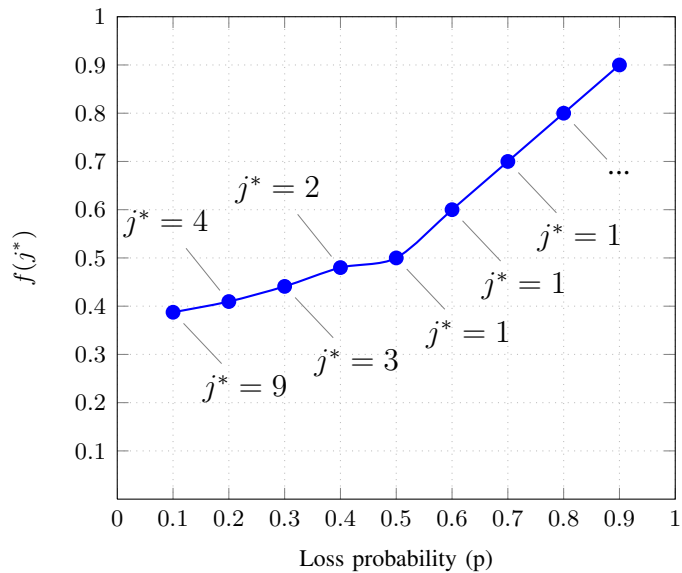
\begin{figure}[t]
\centering
\hspace*{-6mm}
\pgfplotsset{width=9cm, height=8cm}
\tikzset{every mark/.append style={scale=1.2}}
\begin{tikzpicture}
\node[scale=1] {% scale the picture here
\begin{tikzpicture}
	\begin{axis}[
		xlabel= {Loss probability (p)},
		ylabel= {$f(j^*)$}, 
		legend pos=north east,
		xtick={0, .1, .2 , .3, .4, .5, .6, .7, .8, .9, 1},
		ytick={.1,.2,.3,.4,.5, .6, .7, .8, .9, 1},
		tick label style = {font=\footnotesize},
		label style ={font=\footnotesize},
		xmin=0, xmax=1,
		ymin=0, ymax=1,
		grid=major, grid style = dotted,
		legend pos=outer north east, legend cell align=left]

	%\addplot[smooth][color=blue,mark=*,thick] coordinates {
 	%	(.01,.3697) (.05,.3774) (.1,.3874) (.2,.4096) (.3,.4410) (.4,.48)
	%        (.5,.5) (.6,.6) (.7,.7) (.8,.8) (.9,.9) (.95,.95) (.99,.99) 
	%	};
	\addplot[smooth][color=blue,mark=*,thick] coordinates { (.1,.3874) (.2,.4096) (.3,.4410) (.4,.48) (.5,.5) (.6,.6) (.7,.7) (.8,.8) (.9,.9)};
	\node[pin=-70:  {$j^*=9$}] at (axis cs: .1, .3874) {};
	\node[pin=100:  {$j^*=4$}] at (axis cs: .2, .4096) {};
	\node[pin=-70:  {$j^*=3$}] at (axis cs: .3, .4410) {};
	\node[pin=100:  {$j^*=2$}] at (axis cs: .4, .48) {};
	\node[pin=-70:  {$j^*=1$}] at (axis cs: .5, .5) {};
	\node[pin=-70:  {$j^*=1$}] at (axis cs: .6, .6) {};
	\node[pin=-70:  {$j^*=1$}] at (axis cs: .7, .7) {};
	\node[pin=-55:  {...}] at (axis cs: .8, .8) {};
	\end{axis}
\end{tikzpicture}
};
\end{tikzpicture}
\caption{Values of $f(j^*)$ and its corresponding $j^*$. The clique number heavily concentrates around $f(j^*) \times n$, and $j^*$ is the number of packets should be coded together.}
\label{fig:clique-size}
\end{figure}

%\begin{table}[tp]
%\centering
%\begin{tabular} {|c|c|c|c|c|c|c|c|c|c|}
%\hline
%$p$ & 0.1 & 0.2 & 0.3 & 0.4 & 0.5 & 0.6 & 0.7 & 0.8 & 0.9\\
%\hline
%$j^*$ & 9 & 4 &  3 & 2 & 1 & 1 & 1 & 1 & 1\\
%\hline
%$f(j^*)$ & 0.38 & 0.41 & 0.44 & 0.48 & 0.5 & 0.6 & 0.7 & 0.8 & 0.9\\
%\hline
%\end{tabular}
%\caption{\label{table:j-star}Values of $j^*$ and $f(j^*)$ for different loss rates $p$}
%\end{table}

We note that the above concentration result also holds when the number of packets, $m$, is polynomial in the number of user, $n$, \ie, $m = n^d$, for some constant $d > 0$. However, it needs a larger constant $c$ ($c > d\,j$), which means less concentration (as $\delta$ is larger). Apparently, the results do not hold when $m$ is exponential in $n$. However, the cases where $m$ is either linear or polynomial in $n$ are sufficient for practical purposes as in real-time applications, such as \cite{microplay}, $m$ is linear in $n$.

Now let $j^* = \text{argmax} f(j), j^* \in \mathbb{N}$.
%Note that if $j \in \mathbb{R}$, $f(j)$ has a unique maximum at $j = \frac{-1}{\ln q}$.
There may be a set of consecutive values of $j \in \mathbb{N}$ that maximize $f(j)$, in that case, pick $j^*$ to be the smallest value among them. Note that for a constant $p$, $j^*$ and $f(j^*)$ are also constant.
%This suggests that the size of the maximum clique of the graph equals $X_{j^*}$, which is heavily concentrated around $n f(j^*)$.  
\begin{corollary}
For a sufficiently large $n$, with high probability, the maximum clique touches a constant number $j^*$ of columns, where $j^* = \text{argmax} f(j)$.
\end{corollary}

\begin{IEEEproof} Intuitively, this follows from the above result that the size of the maximum clique that touches $j$ columns heavily concentrates around $n f(j)$. In detail, for any constant $j'$ such that $f(j') < f(j^*)$, let $c > \text{max}(j', j^*)+1$. Theorem \ref{thm:boundXj} implies that w.h.p., the size of the maximum clique that touches any $j'$ column is at most 
\[k' = n f(j') + \sqrt{3 c f(j') n \ln n}\,,\] 
and  the size of the maximum clique that touches any $j^*$ column is at least 
\[k'' = n f(j^*) - \sqrt{3 c f(j^*) n \ln n}\,.\]
For a large enough $n$, it is clear that $k' < k''$.
\end{IEEEproof}

Fig. \ref{fig:fj} plots the function $f(j)$ for different values of $p$. This plot shows that (i) for $p>=0.5$, $f(j)$ is a decreasing function, and for $p < 0.5$, $f(j)$ initially increases then decreases, and (ii) $j^*$ increases as $p$ decreases, which suggests the following result: 
\begin{quote}
{\em The number of packets should be coded together increases when the loss rate decreases}.
\end{quote}
%Table \ref{table:j-star} lists the values of $j^*$ and $f(j^*)$ for different loss rates
Fig. \ref{fig:clique-size} plots the values of $f(j^*)$ and the corresponding values of $j^*$.
An important observation from 
%Table \ref{table:j-star} 
Fig. \ref{fig:clique-size} 
is that even when the loss rate is small, the clique size is still high. For instance, when $p=0.1$, we have $j^* = 9$ and $f(j^*) \simeq 0.38$, which means that the optimal coded packet involves coding 9 plain packets together, and this packet will benefit about 38\% of the users.

%-----------------------------------------------------------------
\subsection{Finding a Maximum Clique}
\label{subsec:maximum-clique-random-idnc}

Based on the analysis in the previous section, we propose Max Clique (Algorithm \ref{alg:find-max-clique}) to find a maximum clique of a given random IDNC graph.  Max Clique examines all cliques that touch $j$ columns, for all $j$ combinations of $m$ columns, where $j$ is within a small constant $\delta$ neighborhood of $j^*$. In the case $j^*$ is larger than $m$, $j^*$ is set to $m$ (Line 1), exploiting the fact that for $j < j^*$, $f(j)$ is an increasing function as shown in Fig. \ref{fig:fj}.

{\bf Complexity.} In Max Clique, the for each loop starting at Line 3 runs at most $2 \delta {m \choose {j^* + \delta}}$ times. The for loop starting at Line 5 runs $n$ times. The if condition check at Line 6 examines up to $j^* + \delta$ entries. Thus, the total runtime of Algorithm \ref{alg:find-max-clique} is  at most $2 \delta {m \choose {j^* + \delta}} n (j^* + \delta) = O (n\,m^{j^* + \delta})$, which is polynomial in $n$ when $m$ is linear or polynomial in $n$.

{\bf Optimal Coded Packet.} Given the vertices of the maximum clique output by Max Clique, one can readily compute the optimal instantly decodable packet by XORing the packets whose indices correspond to the packet indices of the output vertices, as indicated in Proposition \ref{proposition:optimal-code-max-clique}. 

% Algorithm to find max clique
\algrenewcommand{\algorithmiccomment}[1]{// {\em #1}}
\algrenewcommand{\algorithmicrequire}{\textbf{Input:}}
\algrenewcommand{\algorithmicensure}{\textbf{Output:}}

\algblockdefx[ForEach]{StartForEach}{EndForEach}%
	[1][]{{\bf for each} #1}%
	{{\bf end}}%

\begin{algorithm}[t]
\begin{algorithmic}[1]

\Require {$p$: loss probability, $n$: number of users, $m$: number of packets, $\matr{A}$: side information matrix of size $n \times m$.}

\Ensure {$\mathcal{I}^*$: vertices of the maximum clique}
\State $j^* \leftarrow \min(m, \text{argmax}_{j \in \mathbb{N}} \, f(j)) $
\State $\mathcal{I}^* = \emptyset$
\StartForEach {combination of $j$ columns out of $m$ columns, where $j \in [j^*-\delta, j^*+\delta]$}
	\State $\mathcal{I} = \emptyset$
	\For {$r = 1 \to n$}
		\If {row $r$ has only one 1 at column $c$}
			\State {Add $(r,c)$ to $\mathcal{I}$}
		\EndIf
	\EndFor
	\If {$| \mathcal{I} | > | \mathcal{I}^* |$}
		\State $\mathcal{I}^* = \mathcal{I}$
	\EndIf
\EndForEach 
\end{algorithmic}
\caption{{\em Max Clique}: Finding the Maximum Clique}
\label{alg:find-max-clique}
\end{algorithm} 

%\textcolor{blue}{Athina-to-Anh: we discussed on skype the assumption m being linear with n. Can you add a paragraph somewhere in this section that you explain what this constraint means, why it is interesting/practical etc.}

%----------------------- Evaluation -----------------------------------
\section{Performance Evaluation}
\label{sec:evaluation}

\subsection{Numerical Evaluation}

\begin{figure*}[tp]
\centering
\subfigure[$n =20$ users, $m=20$ packets]{\includegraphics[width=8cm]{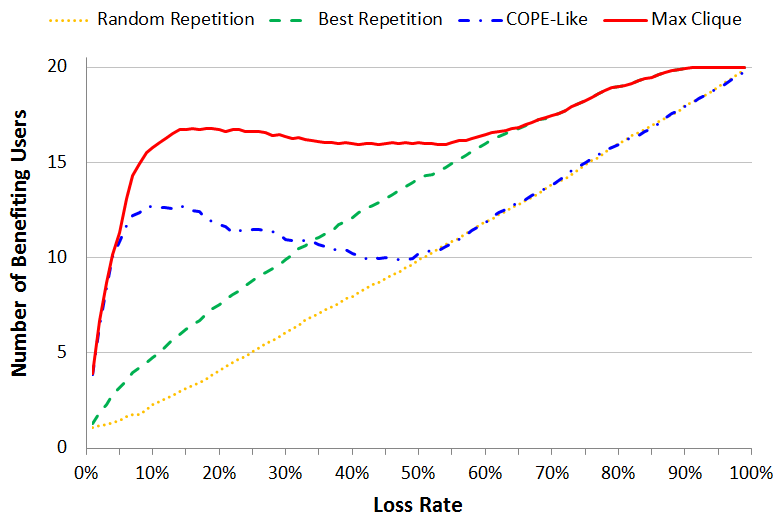}\label{fig:perf1-a}}\hspace{1cm}
\subfigure[$n = 40$ users, $m=20$ packets]{\includegraphics[width=8cm]{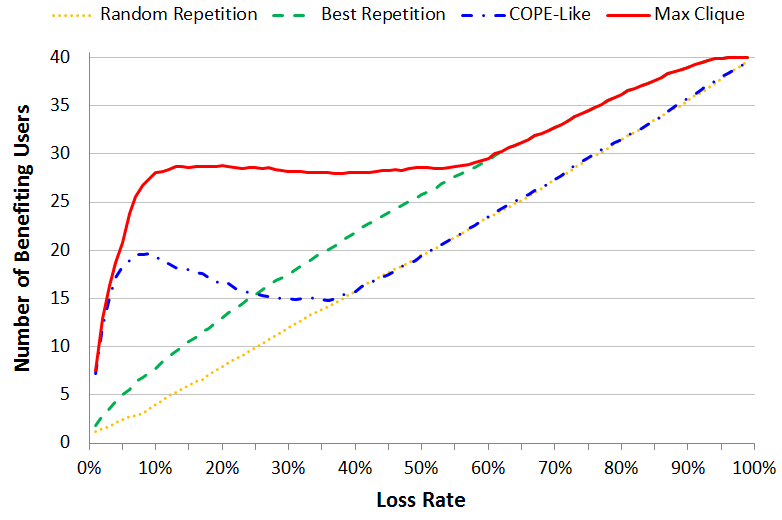}\label{fig:perf1-b}}
\caption{Performance of the proposed Max Clique coding scheme in comparison with those of the Best Repetition and COPE-Like coding schemes.\label{fig:perf1}}
\end{figure*}

In this section, we use simulation to compare the performance of  the proposed Max Clique algorithm (Algorithm \ref{alg:find-max-clique}) against two baselines proposed in \cite{idnc-netcod08}: an optimal repetition-based algorithm, called Best Repetition and a COPE-like greedy-based algorithm. 

The {\em  Best Repetition} algorithm rebroadcasts the plain packet that is wanted by the most number of users. This is inherently the best repetition strategy. The {\em COPE-Like} algorithm goes through all the packets that are still wanted by at least one user in a random order, and it tries to compute a coded packet that is instantly decodable to {\em all} users. In particular, it begins by selecting the first packet of a random permutation, $c = p_1$. It then goes through the rest of the packets one by one. At each step $j$, $j > 1$, it XORs the packet $p_j$ under consideration with $c$: $c = c \oplus p_j$, if the result is still instantly decodable to all users; otherwise, it skips $p_j$. For reference, we also include the Random Repetition algorithm, which resends a random packet that is still wanted by at least one user. 
%This is the simple repetition algorithm proposed in \cite{idnc-netcod08}.

{\flushleft \bf Settings.} For each loss rate ranging from 1\% to 99\%, per 1\% increment, we randomly generate 100 side information matrices. We then run the algorithms on these matrices. For the Max Clique algorithm, we set $\delta$, the neighborhood around $j^*$, to 3. Fig. \ref{fig:perf1} plots the average numbers of beneficiary users as a function of loss rate for the two parameter settings \{$n=20, m=20$\} and \{$n=40, m=20$\}. For clarity, we skip plotting the standard deviations: they are ranging from 0 to 3 for all algorithms. 

{\flushleft \bf Results.} In Fig. \ref{fig:perf1}, we can see that the proposed Max Clique algorithm consistently and significantly outperforms all other algorithms. In particular, for the case \{$n=20, m=20$\}, on average, Max Clique performs 1.3 times better than both the Best Repetition and COPE-Like. For the loss rates between 40\% and 50\%,  Max Clique performs up to 1.6 times better than the COPE-Like algorithm, and for the loss rates between 10\% and 15\%, Max Clique performs up to 3.8 times better than the Best Repetition algorithm. Similar trend but higher improvement, 1.35 times on average and up to 4.5 times, could be observed for the case \{$n=40, m=20$\} in Fig \ref{fig:perf1-b}.

Two interesting observations can be made from Fig. \ref{fig:perf1-a} (and similarly for Fig. \ref{fig:perf1-b}): (i) When the loss rate is larger than a certain threshold (65\% in Fig. \ref{fig:perf1-a}), the performance of Max Clique is similar to that of the Best Repetition, which suggests that Max Clique also tries to select the best uncoded packet. This is because a plain packet now benefits many users due to high loss rate. (ii) When the loss rate is larger than another threshold (50\% in Fig. \ref{fig:perf1-a}), the performance of COPE-Like is similar to that of Random Repetition, which suggests that packets cannot be coded together  while being instantly decodable to all users. This is because when the loss rate is high, given any pair of 2 plain packets, there exists a user who lost both w.h.p.

\subsection{Trace-Based Evaluation}

\begin{figure}[tp]
\centering
\includegraphics[width=8cm]{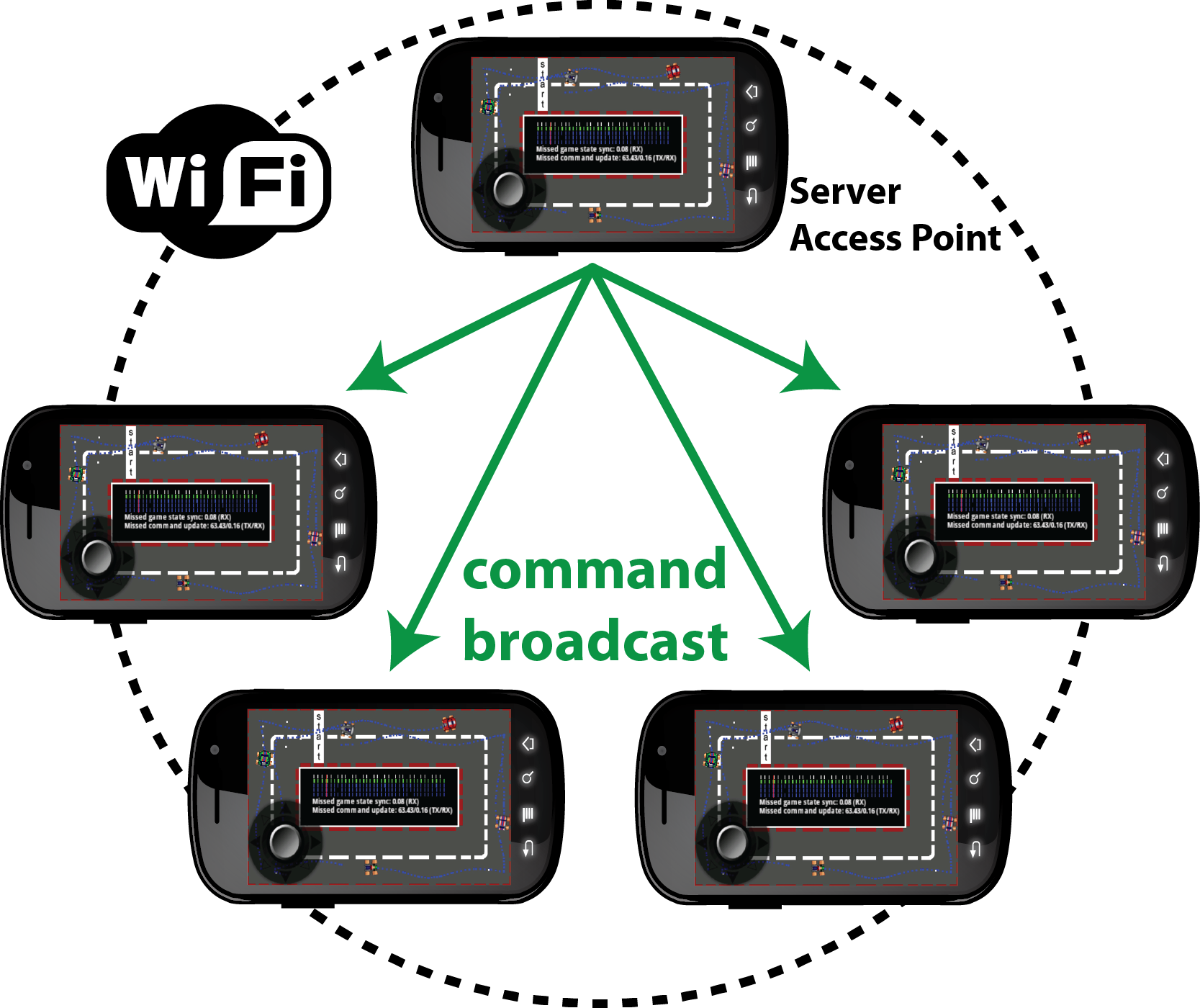}
\vspace*{5pt}
\caption{MicroPlay networking model: One phone acts as the WiFi access point and as the game server. This phone uses WiFi broadcast to disseminate its game commands.}
%\vspace*{5pt}
\label{fig:model}
\end{figure}

\begin{figure*}[tp]
\centering
\subfigure[Reception rate]{\hspace{-0.3cm}\includegraphics[width=9cm]{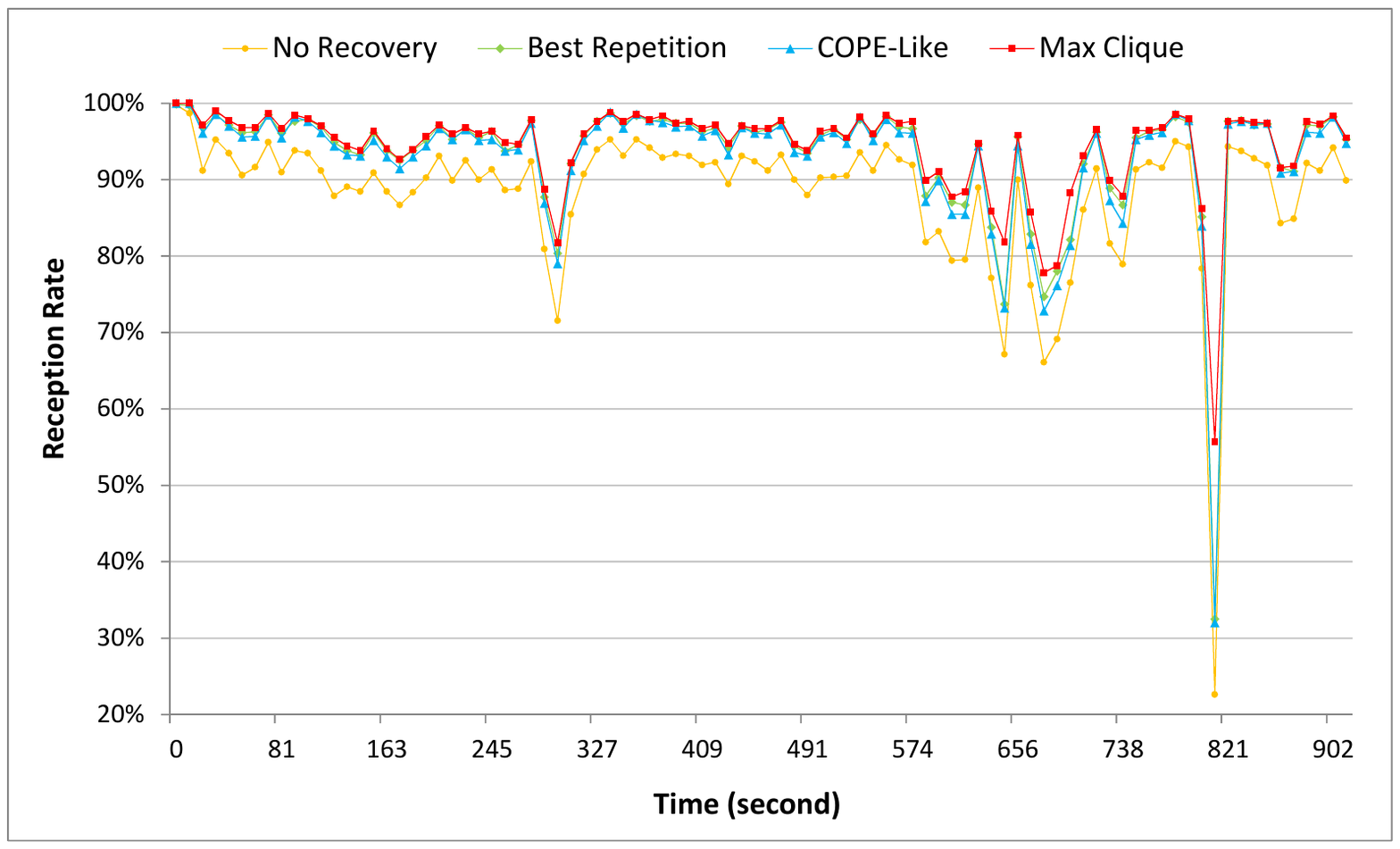}\label{fig:reception}}
\subfigure[Number of beneficiary users]{\includegraphics[width=9.25cm]{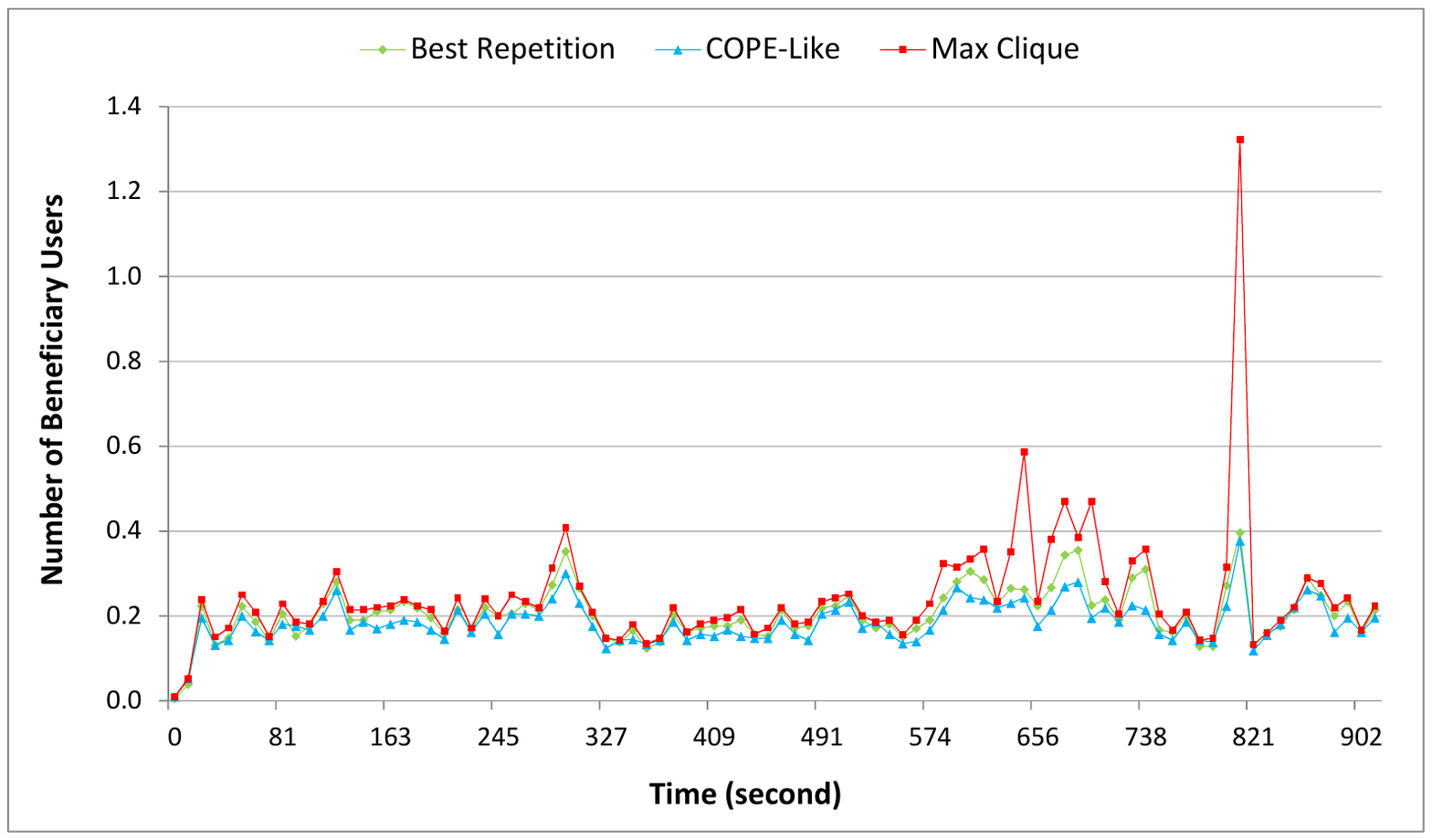}\label{fig:recovery}}
\caption{Trace-based performance of Max Clique in comparison with the baselines when 1 recovery packet is broadcast per 10 packets. The recovery packet is assumed to be received successfully.\label{fig:trace-eval}}
\end{figure*}

In this section, we evaluate the performance of Max Clique in comparison with the baselines, using real network traces of an Android application called Racer \cite{microplay}. Racer is a real-time multi-player racing game implemented on top of a networking framework, called MicroPlay, that we previously developed \cite{microplay}. MicroPlay exploits wireless broadcast to disseminate input commands from one player to the rest in a timely manner to support accurate game rendering and low latency. 

In particular, in Racer, each player's car races around a closed rectangular track and broadcast its movement continuously to the rest of the players. A player uses the broadcast packets to update the positions of the other players' cars. In the context of this work, we examine the packets broadcast by one player, who is acting as the game server and the WiFi access point to the group, depicted in Fig. \ref{fig:model}. This scenario we select for evaluation here, in principle, matches the broadcast scenario that we examined earlier in our analysis in Fig. \ref{fig:broadcast}.

{\flushleft \bf Trace Collection and Description.} We created a Racer game session that has 5 players: 1 server and 4 clients, as shown in Fig. \ref{fig:model}.  The hardware in use consist of 3 Samsung Captivate and 2 Nexus S phones, all running Android OS 2.3 (Gingerbread). The players are scattered  in an on-campus cafeteria, whose area is of sizes approximately 40 x 40 meters. The game session occurs during a busy lunch hour\footnote{We also capture multiple network traces in other hours. We report here the representative traces.}. 

Each packet broadcast by the server has a unique identification number. We implemented a statistics-collection software module within the Racer game client to capture the reception of the packets broadcast by the server. In particular, each client logs the packets it were able to receive and the time it received them. The game session lasted about 15 minutes, and during the game, the server broadcast 19,059 packets, about a packet every 47 ms on average. 

The average reception rate of all 4 clients during the game is shown in Fig. \ref{fig:reception} by the "No Recovery" line. Each point plotted represents the average reception rate of packets broadcast within a 10-second bin. Fig. \ref{fig:reception} shows that the average reception rate of the clients is high: most of the time above 90\%. Nevertheless, there are several instances when the average reception rate drops below 90\%, for example, from second 574 to 738. Also, the average reception rate drops as low as 23\% at second 811. The reception rates are quite similar across the clients. For this reason, we skip reporting the plots of the individual client rates.

{\flushleft \bf Settings.} For each batch of packet of size $B$, we compute a recovery packet using the Best Repetition, COPE-Like, and Max Clique algorithms. This recovery packet is to be broadcast at the end of each batch by the server to recover packet losses at the client. For evaluation purposes, we assume that this packet would be successfully received by all the clients. We then compute the new reception rates at the clients for each recovery scheme.

{\flushleft \bf Results.} Fig. \ref{fig:reception} plots the average reception rate when each of the recovery schemes is used for batch size $B=10$. It could be observed from this figure that Max Clique consistently outperforms the COPE-Like and Best Repetition. In other words, the improvement of the average reception rate is higher when Max Clique is used to compute the recovery packet. 

In more details, Fig. \ref{fig:recovery} plots the number of beneficiary users when each of the recovery scheme is used. Each point plotted is the average over multiple recovery packets within a 10-second bin. Fig. \ref{fig:recovery} shows that the recovery packets computed by Max Clique consistently benefit more users: on average, Max Clique helps 16\% more users than Best Repetition and 26\% more users than COPE-Like. The performance gaps between Max Clique and the baselines are more noticeable when the reception rates are low, \eg, between second 574 and 738, or at second 811, where Max Clique helps 50--250\% more users than the others. 

We also perform similar evaluation for batches of sizes $B=5$ and $B=20$. For $B=5$, the average performance improvement of Max Clique over Best Repetition is 5\% and over COPE-Like is 12\%, which are less than those when $B=10$. This is due to the reduced number of coding opportunities (over just 5 packets). For $B=20$, the average performance improvement of Max Clique over Best Repetition is 12\% and over COPE-Like is 28\%, which are similar to those when $B=10$. This implies that $B=10$ creates sufficient coding opportunities for the loss rates of this set of traces.

Finally, unlike the numerical results reported in the previous section, Fig. \ref{fig:trace-eval} shows that Best Repetition consistently outperforms COPE-Like. This is likely due to the dependency of the packet losses at the clients: a packet lost at a client is likely to be lost at other clients, which implies that re-sending this packet might benefit many clients. This also occurs when $B=5$ and $B=20$.

%---------------------- Conclusion -----------------------------------
\section{Conclusion}
\label{sec:conclusion}
In this paper, we formulate the Real-Time IDNC problem, which seeks to compute a recovery packet that is immediately beneficial to the maximum number of users. Our analysis shows that Real-Time IDNC is NP-Hard. 
%This is shown by first mapping the problem to the problem of finding maximum cliques in IDNC graphs and then providing a reduction from the Exact Cover by 3-Sets problem. 
We then analyze the Random Real-Time IDNC, where each user is assumed to lose every packet with the same probability independently. When the number of packets is linear or polynomial in the number of users, we show that the optimal packet could be computed in polynomial time in the number of users w.h.p., and we provide an explicit algorithm to find the optimal packet.
%We achieve this by providing an algorithm that is capable of finding a maximum clique in a random IDNC graph in polynomial time, which could be of independent interest. 
We evaluate the proposed algorithm numerically as well as experimentally based on real network traces. The results of the evaluation confirm the superior performance of the proposed algorithm.
In the future, we plan to extend this work from a single recovery time slot to a constant number of time slots, corresponding to larger delay tolerance.

%--------------------- Appendix
\appendices
\section{Proof of Lemma 2}
\label{appendix:lemma2}
\begin{IEEEproof}
{\flushleft ($\Leftarrow$)} We first show that a clique in $G^\matr{A}$ maps to an instantly decodable packet, which is uniquely identified by the user and packet indices of the vertices in the clique.

Let $\mathcal{C}$ be a clique in $G^\matr{A}$: $\mathcal{C} = \{v_{i_1, j_1}, \cdots, v_{i_k, j_k}\}$. Without loss of generality, assume $j_1, \cdots, j_k$ are pair-wise distinct, compute $c = p_{j_1} \oplus \cdots \oplus p_{j_k}$. (If $j_{t_1} = j_{t_2} = \cdots = j_{t_n}, n > 1$ then include only $j_{t_1}$ in the XOR.) $c$ is an instantly decodable packet with respect to the set of users $\{u_{i_1}, \cdots, u_{i_k}\}$ because

\squishlist
\item For any user $u_{i_t}$, for some $t \in [1,k]$, the existence of vertex $v_{i_t, j_t}$ indicates that it wants $p_{j_t}$. In the following, we show that $u_{i_t}$ can decode for $p_{j_t}$ immediately upon receiving $c$. Without loss of generality, consider user $u_{i_1}$. It suffices to show that $u_{i_1}$ has all other packets in $c$. To see this, assume otherwise, \ie, assume $u_{i_1}$ does not have packet $p_{j_s}$, for some $s \in [2,k]$ where $p_{j_s} \neq p_{j_1}$. Then there is no edge between $v_{i_1, j_1}$ and $v_{i_s, j_s}$. (contradiction)
\item Each component, $p_{j_t}, t \in [1,k],$ of $c$ is wanted by $u_{i_t}$.
\squishend

{\flushleft ($\Rightarrow$)} We now show that an instantly decodable packet maps to a clique in $G^\matr{A}$, which is uniquely identified by the packets involved and the set of beneficiary users. Let $c^{\mathcal{N}} = p_{j_1} \oplus \cdots \oplus p_{j_k}$ be an instantly decodable packet with respect to the set of user $\mathcal{N}$. Let $p_{j_t}$ be wanted by distinct users $\{u^{j_t}_1, \cdots, u^{j_t}_{n_t}\}$, for some $n_t > 0$. The following set of vertices, $\mathcal{C}$, form a clique in $G^\matr{A}$:
\[ \mathcal{C} = \{ v_{u^{j_1}_1, j_1}, \cdots, v_{u^{j_1}_{n_1}, j_1}, \cdots \cdots, v_{u^{j_k}_1, j_k}, \cdots, v_{u^{j_k}_{n_k}, j_k} \}  \,.\]
We will show that there is an edge between any two vertices in $C$: 
\squishlist
\item For any $t \in [1,k]$, consider any pair $a \neq b, a,b \in [1,n_t]$. There is an edge of type (i) between $v_{u^{j_t}_a, j_t}$ and $v_{u^{j_t}_b, j_t}$ since both $u^{j_t}_a$ and $u^{j_t}_a$ need $p_{j_t}$.
\item For any pair of $s \neq t, s, t \in [1,k]$, consider $a \in [1, n_s]$ and $b \in [1, n_t]$. 
There is an edge of type (ii) between $v_{u^{j_s}_a, j_s}$ 
and $v_{u^{j_t}_b, j_t}$. This is because $u^{j_s}_a$ must have 
$p_{j_t}$ as it can decode for $p_{j_s}$ immediately, and $u^{j_t}_b$ must have 
$p_{j_s}$ as it can decode for $p_{j_t}$ immediately.
\squishend
Finally, it is easy to check that for the above two mappings, one is the reverse mapping of the other.
\end{IEEEproof}

\section{Proof of Lemma 3}
\label{appendix:lemma3}
\begin{IEEEproof}
Let $c^{\mathcal{N}} = p_{j_1} \oplus \cdots \oplus p_{j_k}$ be an instantly decodable packet w.r.t.  the set of user $\mathcal{N}$. Let $p_{j_t}$, $ t \in [1,k]$, benefits distinct users $\{u^{j_t}_1, \cdots, u^{j_t}_{n_t}\}$, for some $n_t > 0$. The following set of vertices, $\mathcal{C}$, forms the clique corresponding to $c^{\mathcal{N}}$:
\[ \mathcal{C} = \{ v_{u^{j_1}_1, j_1}, \cdots, v_{u^{j_1}_{n_1}, j_1}, \cdots \cdots, v_{u^{j_k}_1, j_k}, \cdots, v_{u^{j_k}_{n_k}, j_k} \}  \,.\]

To show $|\mathcal{N}| = |\mathcal{C}|$, it suffices to show that all user indices of vertices in $\mathcal{C}$ are pair-wise distinct. For any pair of $s \neq t$, where $s, t \in [1,k]$, consider any $a \in [1, n_s]$ and any $b \in [1, n_t]$. $u^{j_s}_a \neq u^{j_t}_b$ because otherwise $u^{j_s}_a$ cannot decode for $p^{j_s}$.
\end{IEEEproof}

\section{Proof of Lemma 9 Recurrence}
\label{appendix:lemma9}
\begin{IEEEproof}
It can be shown by induction that 
\begin{align}
B^j_k &= \sum_{i=0}^{j-1} (-1)^i \, {j \choose i} (j-i)^k \,.\label{eqn:cform}
%~ &= j^k - j(j-1)^k + {j \choose 2} (j-2)^k - {j \choose 3} (j-3)^k \cdots \nonumber
\end{align}

In detail, assume that (\ref{eqn:cform}) is true for all indices $1,2, \cdots, j-1$, then following from recurrence $(\ref{recurrence})$,
\begin{align*}
B^j_k &= j^k - {j \choose 1} B^{j-1}_k - {j \choose 2} B^{j-2}_k \cdots - {j \choose j-1} B^{1}_k\\
~ &= j^k - {j \choose 1} \sum_{i=0}^{j-2} (-1)^i \, {{j-1} \choose i} (j-1-i)^k \\
~ &\quad~\quad- {j \choose 2} \sum_{i=0}^{j-3} (-1)^i \, {{j-2} \choose i} (j-2-i)^k\\
~ &\quad~\quad - \cdots \\
~ &\quad~\quad - {j \choose j-1} \cdot 1 \,.
\end{align*}

Now, for any $t \in [1, j-1]$, the coefficient of $(j-t)^k$ is 
\begin{align*}
\sum_{i=1}^t {j \choose i} {j-i \choose {t - i} } (-1)^{t - i+1}\,.
\end{align*}

Thus, it suffices to show that, for any $t \in [1, j-1]$, 
\begin{align*}
\sum_{i=1}^t {j \choose i} {j-i \choose {t - i} } (-1)^{t - i+1} &= (-1)^t {j \choose t} \,.
\end{align*}

The above equation holds iff
\[ \sum_{i=1}^t \frac{ (-1)^{t-i+1} }{ i! \, (t-i)!} = \frac{(-1)^t}{t!}\,.\]

Or, equivalently
\[ \sum_{i=1}^t {t \choose i} (-1)^{t-i}  = (-1)^{t+1} \,.\]

The LHS of the above equation equals 
\[ - (-1)^{t} + \sum_{i=0}^t {t \choose i} (-1)^{t-i}  = (-1)^{t+1}\,, \]
where the last ``='' follows from the binomial theorem (for $a=1, b=-1$).
\end{IEEEproof}

\section{Proof of Lemma 10}
\label{appendix:lemma10}
\begin{IEEEproof}
Denote $Z_{\mathcal{C}_j}$ and $X_{\mathcal{C}_j}$ by $Z$ and $X$, respectively. Applying Chernoff's bound on the Binomial distributed variable $Z$, we have
\[ \text{Pr} [ \,|\, Z  - \mu \,|\, \geq \mu \delta ] \leq 2 \, \text{exp}(- \frac{\mu \delta^2}{3}) = \frac{2}{n^c} \,. \]

Now, 
\begin{align*}
~&\text{Pr}[ \,|\, X - \mu  \,|\, \geq \mu \delta ]  \\
~&= \sum_{k=1}^n \text{Pr}[ \,|\, X - \mu  \,|\, \geq \mu \delta \,|\, Z = k ] \cdot \text{Pr}[Z = k] \\
~&\leq \sum_{k=1}^{\mu-\mu \delta} \text{Pr}[ Z = k ]  + \sum_{k=\mu+\mu \delta}^{n} \text{Pr}[ Z = k ]\\
~&\quad\quad + \sum_{k=\mu-\mu \delta + 1}^{\mu+\mu \delta - 1} \text{Pr}[ \,|\, X - \mu  \,|\, \geq \mu \delta \,|\, Z = k ] \\
~&\leq \frac{2}{n^{c}} + \sum_{k=\mu - \mu \delta + 1}^{\mu+\mu \delta -1 }  \text{Pr}[ \,|\, X - \mu  \,|\, \geq \mu \delta \,|\, Z = k ]\\
%~&= \frac{2}{n^{c}} + \sum_{k=\mu - \mu \delta + 1}^{\mu+\mu \delta - 1}  \text{Pr}[ \,|\, X - \mu  \,|\, \geq \mu \delta \,|\, Z = k, X \neq Z ] \cdot \text{Pr}[X \neq Z]\\
%~&\quad\quad\quad\quad\quad\quad +   \text{Pr}[ \,|\, X - \mu  \,|\, \geq \mu \delta \,|\, Z = k, X = Z ] \cdot \text{Pr}[X = Z]\\
%~&\leq \frac{2}{n^{c}} + (2\mu \delta - 2) \, j(1-1/j)^{\mu - \mu \delta} \quad \text{(from Lemma \ref{lemma:XZ} )} \\
%~&\quad\quad + \sum_{k=\mu - \mu \delta + 1}^{\mu+\mu \delta - 1}  \text{Pr}[ \,|\, Z - \mu  \,|\, \geq \mu \delta \,|\, Z = k ] \\
%~&\leq \frac{2}{n^{c}} + 2\mu \delta \,  j(1-1/j)^{\mu - \mu \delta} + 0\,.
\end{align*}

For $Z \in [\mu - \mu \delta +1, \mu + \mu \delta - 1]$, we have
\begin{align*}
~&~\text{Pr}[ \,|\, X - \mu  \,|\, \geq \mu \delta ]\\
~&~=\text{Pr}[ \,|\, X - \mu  \,|\, \geq \mu \delta, X=Z ] \cdot \text{Pr}[X = Z] \\
~&~\quad+ \text{Pr}[ \,|\, X - \mu  \,|\, \geq \mu \delta, X \neq Z ] \cdot \text{Pr}[X \neq Z]\\
&~\leq \text{Pr}[ \,|\, X - \mu  \,|\, \geq \mu \delta, X=Z ] + \text{Pr}[X \neq Z]\\
&~= 0 + \text{Pr}[X \neq Z]\\
&~\leq j(1-1/j)^{\mu - \mu \delta} \quad \text{(from Lemma \ref{lemma:XZ} )}
\end{align*}

Thus, 
\[
\text{Pr}[ \,|\, X - \mu  \,|\, \geq \mu \delta ] \leq \frac{2}{n^{c}} + 2\mu \delta \,  j(1-1/j)^{\mu - \mu \delta}\,.
\]
\end{IEEEproof}

%--------------------- References

%\textcolor{blue}{Athina: I think you have more references tahn actually cited in the paper. Especially after I commented out parts of related work. Remove some references. Also, condense short lines.}

\balance

%--------------- Conclusion ----------------------------------------
%\input{conclusion.tex}

%----------------------------------------------------------------------------------------------------------------- ACKNOWLEDGMENT
%\section{Acknowledgment}
%This work has been supported by an NSF CAREER award (0747110) and by AFOSR MURI  (FA9550-09-1-0643).

% -------------- Bibliography --------------------------------------------------
%\bibliographystyle{IEEEtran}
%\bibliography{NetworkCodingSecurity}
%\input{netcod2012.bbl}

% -------------- Appendix --------------------------------------------------
%\input{appendix.tex}

\end{document}